\documentclass[%
twocolumn,
amsmath,amssymb,
aps, physrev,
]{revtex4-2}
\usepackage{enumitem}
\usepackage{graphicx}
\usepackage{dcolumn}
\usepackage{bm}
\usepackage{amsmath}
\usepackage{braket}
\usepackage{booktabs}
\usepackage{rotating}
\usepackage[colorlinks=true,allcolors=blue]{hyperref}
\usepackage{tikz}
\usepackage{xcolor}

\usetikzlibrary{arrows.meta,decorations.pathreplacing}
\usepackage{cancel}
\usepackage{soul}

\begin{document}


\title{\textbf{Coherence-mediated quantum thermometry in a hybrid circuit-QED architecture} }%

\author{Shaojiang Zhu}
\email{Contact author: szhu26@fnal.gov}
\affiliation{Superconducting Quantum Materials and Systems Center, Fermi National Accelerator Laboratory, Batavia, IL 60510, USA}%

\author{Xinyuan You}
\affiliation{Superconducting Quantum Materials and Systems Center, Fermi National Accelerator Laboratory, Batavia, IL 60510, USA}%

\author{Alexander Romanenko}
\affiliation{Superconducting Quantum Materials and Systems Center, Fermi National Accelerator Laboratory, Batavia, IL 60510, USA}%

\author{Anna Grassellino}
\affiliation{Superconducting Quantum Materials and Systems Center, Fermi National Accelerator Laboratory, Batavia, IL 60510, USA}%


\begin{abstract}

Quantum thermometry plays a critical role in the development of low-temperature sensors and quantum information platforms.
In this work, we propose and analyze a hybrid circuit quantum electrodynamics architecture in which a superconducting qubit is dispersively coupled to two distinct bosonic modes: one initialized in a weak coherent state as a phase reference and information buffer and the other coupled to a thermal environment.
We show that the qubit serves as a sensitive readout of the probe mode, mapping the interplay between thermal and coherent photon-number fluctuations onto measurable dephasing.
This coherence-mediated mechanism enables improved sensitivity to thermal energy fluctuations in the sub-millikelvin regime through Ramsey interferometry.
We derive analytic expressions for the probe coherence envelope, compute the quantum Fisher information for temperature estimation, and demonstrate numerically that the presence of a coherent reference enhances the qubit's sensitivity to small changes in thermal photon occupancy.
Our results establish a coherence-enabled approach to thermometry and provide a scalable platform for future calorimetric sensing in high-energy physics and quantum metrology.

\end{abstract}

\maketitle

\section{Introduction}
Quantum thermometry has emerged as a critical capability in modern quantum science and technology. 
It plays a central role in the calibration of cryogenic environments, the characterization of quantum devices and sensors, and the detection of weak energy events in low-temperature physics~\cite{clerk2010introduction, de2019quantum, stace2010quantum, mehboudi2019thermometry}. 
Beyond practical applications, precise temperature measurement also provides a platform for exploring fundamental thermodynamic limits in the quantum regime. 
Superconducting quantum circuits, particularly those operating in the microwave domain, offer a versatile and scalable platform for implementing such thermometric protocols~\cite{devoret2004superconducting, krantz2019quantum}. 
In these systems, qubits serve as exquisitely sensitive probes of their electromagnetic environment and can detect thermal photon populations through their decoherence dynamics~\cite{jevtic2015single, scigliuzzo2020primary, sultanov2021protocol, sharafiev2025leveraging}.

The most common approach to qubit-based thermometry exploits the dispersive interaction between a qubit and a thermalized cavity mode. 
In the dispersive regime, photon-number fluctuations induce pure dephasing of the qubit, which can be monitored via Ramsey sequences~\cite{serniak2018hot, bultink2020protecting}. 
The decay of qubit coherence directly encodes the thermal occupancy $\bar{n}_{th}$ of the mode and thereby its effective temperature. 
This single-mode dephasing protocol has been demonstrated in a variety of circuit quantum electrodynamics (cQED) architectures~\cite{lvov2025thermometry,paris2009quantum,brunelli2011qubit,yan2018distinguishing, medford2013self, mallet2011quantum} and is particularly effective when $\bar{n}_{th} \gtrsim 1$. 
However, in the ultra-low-temperature limit $\bar{n}_{th} \ll 1$, corresponding to mode temperatures below $50$~mK at GHz frequencies, the variance of Bose–Einstein photon statistics becomes exponentially suppressed, and the sensitivity of this method is fundamentally limited by the vanishing signal-to-noise ratio of thermal fluctuations.
Beyond qubit-based approaches, a range of established cryogenic thermometry techniques, such as Coulomb-blockade thermometers and proximity-based superconducting thermometers, provide robust calibration standards and performance across complementary temperature ranges~\cite{saira2016dispersive,wang2018fast,gumucs2023calorimetry}.

Here, we introduce the coherence-mediated quantum thermometry (CMQT), a two-mode quantum thermometry protocol that overcomes this low-temperature sensitivity floor by leveraging coherent-thermal interference. 
We consider a hybrid cQED architecture in which a superconducting qubit is dispersively coupled to two bosonic modes: a high-$Q$ 3D cavity initialized in a weak coherent state, and a low-$Q$ planar resonator coupled to a thermal environment. 
The coherent mode acts as a stable phase reference, while the thermal absorber mode functions as the sensing arm. 
The qubit, coupled to the probe cavity, serves as a phase-sensitive readout whose Ramsey phase acquires a stochastic shift determined by the joint photon-number fluctuations of both modes. 
As a result, the qubit's coherence decay directly reflects the interference between their photon statistics. 
In this framing, the key capability of CMQT is the separation of roles: the absorber is engineered for strong and well-characterized thermalization, the probe acts as a phase-coherent information buffer, and the qubit is only engaged during a short mapping window. This modularity enables independent optimization of the thermal coupling and bandwidth, the storage time and technical noise filtering, and the qubit coherence and readout fidelity, a set of trade-offs that is difficult to realize when the qubit simultaneously serves as sensor, memory, and readout.

This architecture realizes an interferometric thermometry scheme in which small thermal signals are amplified by embedding them in the phase dynamics of a coherent reference. 
We analytically derive the coherence envelope of the probe mode as a function of interaction time, thermal photon number, and coherent amplitude, and show how it is faithfully mapped onto qubit Ramsey dephasing through dispersive readout. 
The resulting dephasing dynamics depend not only on the variances of each mode individually but also on their interference, enabling thermal signal amplification through coherent--thermal correlations. 
To quantify the resulting sensitivity, we calculate the quantum Fisher information (QFI) associated with qubit coherence measurements and identify optimal working points that maximize temperature resolution~\cite{paris2009quantum, demkowicz2015quantum, liu2020quantum, zhong2013fisher}. 
This added hybrid complexity is motivated by the ultra-low-occupation regime $\bar n_a\!\ll\!1$, where conventional single-mode dispersive thermometry exhibits a parametrically small temperature responsivity.

\begin{figure}[b!]
\centering
\includegraphics[width=0.9\columnwidth, trim={4.8cm 1.0cm 4.7cm 5.0cm}, clip]{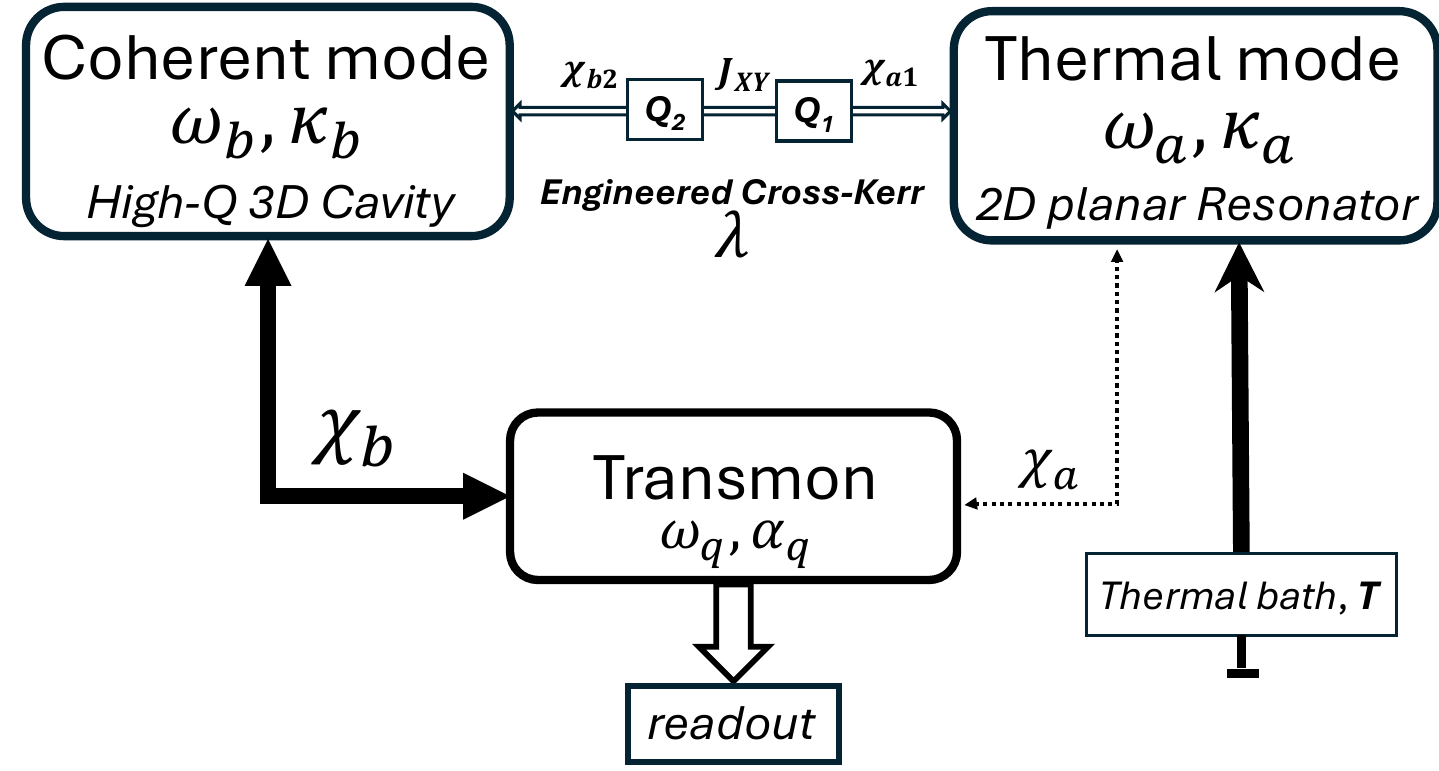}
\caption{
Schematic of the proposed thermometry architecture. 
A low-$Q$ thermal resonator $\hat a$ is equilibrated with a bath at temperature $T$. 
Its photon-number fluctuations are imprinted to a high-$Q$ 3D cavity probe mode $\hat b$ via an engineered cross-Kerr interaction $\lambda$ implemented through two intermediate qubits ($Q_1$, $Q_2$). 
A readout transmon qubit couples dispersively to the probe with strength $\chi_b$ for phase mapping, while any residual coupling $\chi_a$ to the thermal mode introduces parasitic qubit dephasing. 
The temperature-dependent phase statistics encoded in the probe can be extracted either through a qubit Ramsey measurement or through direct heterodyne detection. 
Here $\omega$ denotes the mode frequencies, $\kappa$ their linewidths, and $\alpha_q$ the qubit anharmonicity.
}
\label{fig:schematic}
\end{figure}

Numerical simulations based on exact unitary interaction validate the analytic model and show that sub-millikelvin sensitivity is, in principle, achievable with realistic circuit parameters. 
The proposed architecture is compatible with existing cQED platforms, including hybrid designs that combine planar resonators and qubits with 3D cavities~\cite{axline2016architecture, ganjam2024surpassing, kim2025ultracoherent}, and does not require photon counting or non-Gaussian state preparation. 
These features make it a scalable and noise-resilient foundation for thermometry at ultra-low energies. 
Potential applications range from on-chip quantum diagnostics~\cite{gasparinetti2015fast} and cryogenic calorimetry~\cite{pretzl2000cryogenic} to thermal signal amplification in rare-event detection platforms such as axion or hidden-photon dark matter searches~\cite{brubaker2017first, malnou2019squeezed}. 
Thus, these results establish coherent–thermal interferometry as a promising route toward quantum-limited thermometric sensing in superconducting circuits. 
We note practical constraints that bound performance in realistic devices, including maintaining dispersive operation, calibrating AC-Stark/Kerr drifts and drive-induced leakage under coherent drives, characterizing non-thermal absorber noise, and managing long averaging times at ultra-low $T$ with minimal per-shot backaction. 
Instead, in this work we identify the core CMQT operating point and its regime of advantage using a minimal, transparent model that isolates the essential temperature-to-phase transduction mechanism.

This article is organized as follows. Section~\ref{sec:theory} introduces the theoretical model and hybrid circuit-QED architecture, derives the thermal-to-coherent mapping mechanism, and outlines two operating strategies suited to different measurement protocols and temperature ranges. Section~\ref{sec:experiment} details an experimentally realistic implementation, including mode realization, the engineered cross-Kerr interaction, expected qubit decoherence, and strategies for mitigating excess noise. Section~\ref{sec:comparison} benchmarks CMQT against a conventional qubit-only thermometer in which the qubit acts as both sensor and memory, emphasizing the resulting preservation of qubit coherence and the associated sensitivity improvement. Section~\ref{sec:summary} summarizes the main results and discusses practical limitations and outlook. Further technical derivations are collected in the Appendices.

\section{Theory: Thermal-to-Coherent Mapping}\label{sec:theory}

Quantum thermometry aims to extract information about temperature from quantum systems with minimal disturbance and maximal sensitivity. 
In this work, we consider a hybrid bosonic system where a thermal field is coupled with a coherent reference field. 
This interaction, acting as a noise-to-phase transduction mechanism, enables temperature-dependent fluctuations in the thermal field to be imprinted onto a well-controlled coherent mode, thereby amplifying otherwise weak thermal signals and allowing them to be efficiently detected by monitoring the qubit’s coherence. 
Below we present a detailed theoretical framework for this thermal-to-coherent mapping mechanism.

\subsection{System Model}

As shown in Fig.~\ref{fig:schematic}, we consider a hybrid system comprising a transmon qubit coupled to two bosonic modes:

Mode $\hat{a}$: a thermal field at temperature $T$, initialized in a thermal state $\rho_{\mathrm{th}}(T)$ with mean photon number $\bar{n}_a$. 
A 2D planar resonator, such as a CPW or lumped-element LC resonator, can be used to realize this mode.

Mode $\hat{b}$: a coherent reference field prepared in a coherent state $|\alpha\rangle$ with amplitude $\alpha$. 
This mode can be implemented using a high-quality 3D cavity. 

A strong dispersive coupling $\chi_b$ enables the transmon qubit to map any temperature-dependent phase shift accumulated in the coherent probe mode, whereas the parasitic coupling $\chi_a$ to the thermal mode introduces additional dephasing in the qubit.

In addition, a nonlinear cross-Kerr interaction mediates a joint coupling between the thermal and coherent modes. 
The total interaction Hamiltonian is given by
\begin{equation}
    \hat{H}_{\mathrm{int}} = \hat{\sigma}_z \left( \chi_a \hat{n}_a + \chi_b \hat{n}_b \right) + \lambda \hat{n}_a \hat{n}_b,
    \label{eq:Hamiltonian}
\end{equation}
where $\hat{n}_{a(b)} = \hat{a}^\dagger \hat{a}$ ($\hat{b}^\dagger \hat{b}$) are the photon number operators, $\chi_a$ and $\chi_b$ are dispersive coupling strengths between the qubit and each mode, and $\lambda$ is the cross-Kerr strength. 
The first two terms represent standard dispersive phase shifts proportional to photon number, while the final term provides a direct photon-number–dependent coupling between the two modes. 
Here $\lambda$ arises as an effective interaction mediated by virtual transitions through multiple qubit pathways~\cite{hu2011cross,kounalakis2018tuneable,aoki2024control}. 
In this work, we intentionally decouple the thermal field from the qubit so $\chi_a$ is negligible. 
This avoids qubit decoherence due to direct thermal contact, while preserving thermal sensitivity via the indirect $\lambda$-mediated path.

\subsection{Lamb-Like Shift: Thermal-to-Coherent Mapping}\label{sec:lamb}

The cross-Kerr interaction, $\lambda \hat{n}_a \hat{n}_b$, couples the thermal absorber $a$ to the coherent probe $b$, such that thermal photon-number fluctuations induce frequency shifts in the coherent mode~\cite{blais2021circuit, schuster2005ac}. 
While the traditional Lamb shift arises as a second-order correction to a quantum system's energy levels due to vacuum fluctuations, the present scenario exhibits a reversed structure: it is the coherent field, rather than the qubit, that experiences environment-induced shifts. 
Here, the thermal mode acts as an effective fluctuating environment for the probe.

From the perspective of the coherent mode, the effective Hamiltonian reads
\begin{equation}
    \hat{H}_b^{\mathrm{eff}} = \left( \omega_b + \lambda \hat{n}_a \right) \hat{n}_b,
\end{equation}
where the thermal occupation $\hat{n}_a$ follows Bose-Einstein statistics:
\(
\bar{n}_a(T) = 1/[\text{exp}(\hbar \omega_a / k_B T) - 1].
\)
Due to the probabilistic nature of thermal fluctuations, each realization of $n_a$ photons shifts the frequency of the coherent mode by $\lambda n_a$, resulting in a distribution of frequency shifts centered at $\omega_b + \lambda \bar{n}_a$.

This random frequency shift leads to stochastic phase accumulation. 
For a single realization with $n_a$ photons, the phase shift after the absorber-probe interaction time $\tau$ is deterministic: $\phi_b(\tau, T) = \lambda n_a \tau$. 
Across the thermal ensemble, however, this becomes a random variable:
\(
\phi_b = \int_0^\tau \delta\omega_b(t) \, dt,
\)
where $\delta\omega_b(t)$ captures the instantaneous fluctuation in frequency due to the thermal photon number. 
This process defines a pure dephasing channel for the coherent state: it causes a gradual loss of phase coherence without energy dissipation.

Ensemble-averaging over many thermal realizations transforms the initial pure coherent state into a statistical mixture of phase-rotated states. 
As a result, the coherent mode no longer exhibits a well-defined global phase but instead acquires a broadened phase distribution. 
This dephasing is observable as a decay in the coherent amplitude, and it plays a central role in \textit{Strategy~1}, where we model the coherence envelope as a temperature-sensitive observable.

Importantly, while the original Lamb shift arises as a second-order effect (\( \propto \lambda^2 \)), the thermally induced frequency shift in our model is a first-order effect in \( \lambda \), but becomes effectively stochastic due to the underlying thermal noise. 
The resulting dephasing thus encodes temperature-dependent fluctuations into the probe's coherence dynamics.

\subsection{Strategy 1: Coherence-mediated (CM) envelope and QFI}

\begin{figure}[b!]
\centering
\includegraphics[width=0.95\columnwidth]{CM-QFI.pdf}
\caption{
Simulated coherence-based QFI, \(\mathcal{F}_C\), and corresponding temperature sensitivity, \(\delta T\), for a thermal mode with frequency \(\omega_a/2\pi = 1~\mathrm{GHz}\), probe amplitude \(\alpha = 1.0\), and cross-Kerr coupling \(\lambda/2\pi = 50~\mathrm{kHz}\).
(a) Heatmap of \(\mathcal{F}_C\) as a function of thermal bath temperature \(T\) and interaction time \(\tau\). 
The white dashed curve denotes the optimal interaction time \(\tau_{\mathrm{opt}}(T)\) that maximizes \(\mathcal{F}_C\) at each temperature.
(b) Minimum detectable temperature change \(\delta T_{\mathrm{min}} = 1/\sqrt{\nu \mathcal{F}_C}\) (red), evaluated along \(\tau_{\mathrm{opt}}(T)\) (green). 
A sensitivity of approximately \(\delta T = 60~\mu\mathrm{K}\) is achieved at \(T = 10~\mathrm{mK}\) with \(\tau_{\mathrm{opt}} \approx 25~\mu\mathrm{s}\) for \(\nu = 10^4\) repetitions.
(c) QFI as a function of interaction time \(\tau\) at fixed \(T = 10~\mathrm{mK}\) for two different cross-Kerr strengths \(\lambda\).  
While the maximum attainable QFI is independent of \(\lambda\), the optimal $\tau$ decreases for larger \(\lambda\), reflecting faster phase accumulation.
}
\label{fig:c-qfi}
\end{figure}

After interacting with the thermal mode via a cross-Kerr coupling, the coherent probe acquires temperature-dependent phase noise. 
This dephased coherent state is interrogated by a dispersively coupled qubit using a Ramsey sequence~\cite{clerk2007using, sears2012photon}. 
The qubit’s coherence reflects the fluctuating probe phase: \(|\psi_q(\tau)\rangle = \tfrac{1}{\sqrt{2}} \left(|0\rangle + e^{i \phi_b} |1\rangle\right),\) where the accumulated phase \( \phi_b = \lambda n_a \tau\) originates from thermal fluctuations and is therefore temperature dependent.

We model the probe mode as undergoing random phase kicks induced by photon-number fluctuations in the thermal mode under the quasi-static condition ($\kappa_a \tau < 1$). 
Since $\phi_b \propto n_a$, its statistics are inherited from the Bose-Einstein distribution of $n_a$, with mean $\bar{n}_a$ and variance $\mathrm{Var}(n_a) = \bar{n}_a(\bar{n}_a+1)$. 
Accordingly, the probe phase has mean $\langle \phi_b \rangle = \lambda \tau \bar{n}_a$ and variance
\begin{equation}
\sigma_\phi^2 = \mathrm{Var}(\phi_b) = (\lambda \tau)^2 \bar{n}_a(\bar{n}_a+1).
\end{equation}
In the low-temperature limit ($\bar{n}_a \ll 1$), the variance reduces to $\sigma_\phi^2 \approx (\lambda \tau)^2 \bar{n}_a$. 

Although $n_a$ is not Gaussian distributed, the accumulated phase noise results from many small, independent fluctuations and can be approximated by Gaussian diffusion (central limit theorem)~\cite{clerk2010introduction, gambetta2006qubit}. 
The probe is thus described as a phase-averaged mixture,
\begin{equation}
\rho_b(T) = \int d\phi \, \tfrac{e^{-\phi^2/2\sigma_\phi^2}}{\sqrt{2\pi\sigma_\phi^2}} \,
|\alpha e^{i\phi}\rangle\langle \alpha e^{i\phi}|,
\end{equation}
which captures the temperature-dependent dephasing of the probe field. 
This Gaussian phase-diffusion picture is directly analogous to qubit dephasing from a dispersively coupled thermal cavity, with coherence loss governed by the same $\bar{n}_a(\bar{n}_a+1)$ scaling~\cite{gambetta2006qubit, schuster2007resolving, catelani2011quasiparticle, fabre2020modes}.

The probe coherence envelope under Gaussian phase diffusion is
\begin{equation}
\label{eq:C_Gamma}
C(\tau, T) = \exp\left[ -2|\alpha|^2 \left( 1 - e^{-\Gamma_\phi} \right) \right],
\end{equation}
with the effective dephasing rate \(\Gamma_\phi = \sigma_\phi^2/2\) (Appendix~\ref{app:C_envelope}).

This expression reflects the progressive decoherence of a coherent state subject to Gaussian phase fluctuations with temperature-dependent variance, and it is experimentally accessible via Ramsey measurements of the qubit. 
At low temperatures ($\bar{n}_a \to 0$), dephasing vanishes and $C(\tau, T) \to 1$. 
At high temperatures, the envelope saturates at $\exp(-2 \alpha^2)$, reflecting complete phase scrambling~\cite{serniak2018hot}. 
The nonlinear dependence on both $\alpha$ and $\bar{n}_a$ enables strong sensitivity to thermal fluctuations even in the sub-photon regime.

To quantify thermometric performance, we compute the quantum Fisher information (QFI) associated with the measurable signal \( C \):
\(
\mathcal{F}_C(T) = |\partial_T C|^2 / (1-C^2),
\)
with $\Phi(T)$ constant (Appendix~\ref{app:QFI_definition}). Substituting $C(\tau,T)$ gives
\begin{equation}\label{eq:qfi_cm}
\mathcal{F}_C(T) = \frac{\left[ \alpha^2 e^{-\Gamma_{\phi}} (\lambda \tau)^2 \, (1+2\bar{n}_a)\, \partial_T \bar n_a \right]^2 C^2}{1 - C^2},
\end{equation}
where \(\partial_T \bar n_a = \hbar \omega_a  \!\left[ \bar n_a (\bar n_a +1) \right]  /k_B T^2\).
Eq.~\ref{eq:qfi_cm} establishes the fundamental precision bound of the CM thermometer~\cite{paris2009quantum,liu2020quantum,zhong2013fisher, demkowicz2015quantum, correa2015individual}.

Physically, the probe converts temperature-dependent photon-number fluctuations into measurable phase diffusion. 
In this CM model, the QFI increases with $\alpha^2$ but is simultaneously limited by visibility decay in the denominator. 
In the weak-dephasing regime ($\Gamma_\phi \ll 1$), one finds
\(
\mathcal{F}_C \;\approx\; \alpha^2 (\lambda\tau)^2 \,\bar n_a \left(\hbar \omega_a / k_B T^2 \right)^{2}.
\)
Thus, the coherent-probe scheme provides a tunable transduction “gain” via $\alpha$ (and $\tau$), but this gain is bounded in practice by visibility decay, available probe power, bandwidth, and backaction. 
We therefore interpret the improvement as a resource-assisted sensitivity, which is achieved by allocating probe photons to strengthen the temperature-to-phase conversion, rather than as an unlimited amplification.

Fig.~\ref{fig:c-qfi}(a) shows a simulated QFI, $\mathcal{F}_C$, heatmap, highlighting the optimal interaction time $\tau(T)$ shown as white dashed line for the maximized QFI at each temperature.

We define the temperature sensitivity as \(\delta T_{\mathrm{min}} = 1 / \sqrt{\nu \mathcal{F}_C(T)},\) with $\nu$ the number of independent repetitions.
Fig.~\ref{fig:c-qfi}(b) shows that the thermometer can achieve a sensitivity $\delta T \sim 60~\mu\mathrm{K}$ at $T = 10~\mathrm{mK}$, with an optimal interaction time around $\tau \sim 25~\mu\mathrm{s}$ and measurement repetition $\nu = 10^4$.
Such timescales are readily supported by high-Q 3D cavities, which exhibit photon lifetimes up to millisecond range~\cite{reagor2013reaching, romanenko2020three, kim2025ultracoherent}.

The calculated QFI as a function of $\tau$ for different cross-Kerr strengths is shown in Fig.~\ref{fig:c-qfi}(c), with fixed temperature $T$ = 10 mK. 
Since the qubit remains in its ground state during the sensing window $\tau$ and is only activated briefly afterward to map out the probe coherence (Appendix.~\ref{app:mapping}), this scheme provides a useful tuning knob to balance the cross-Kerr strength $\lambda$ and the interaction time $\tau$. 
A longer sensing window allows operation with a smaller $\lambda$, which in turn suppresses probe-induced backaction on the thermal absorber and permits a larger probe amplitude $\alpha$.
Conversely, a stronger $\lambda$ enables faster sensing with shorter $\tau$, while maintaining moderate backaction by reducing $\alpha$~\cite{rigetti2012superconducting}.

\subsection{Strategy 2: Probe Phase-Shift (PS) Tracking}

\begin{figure}[b!]
\includegraphics[width=0.95\columnwidth]{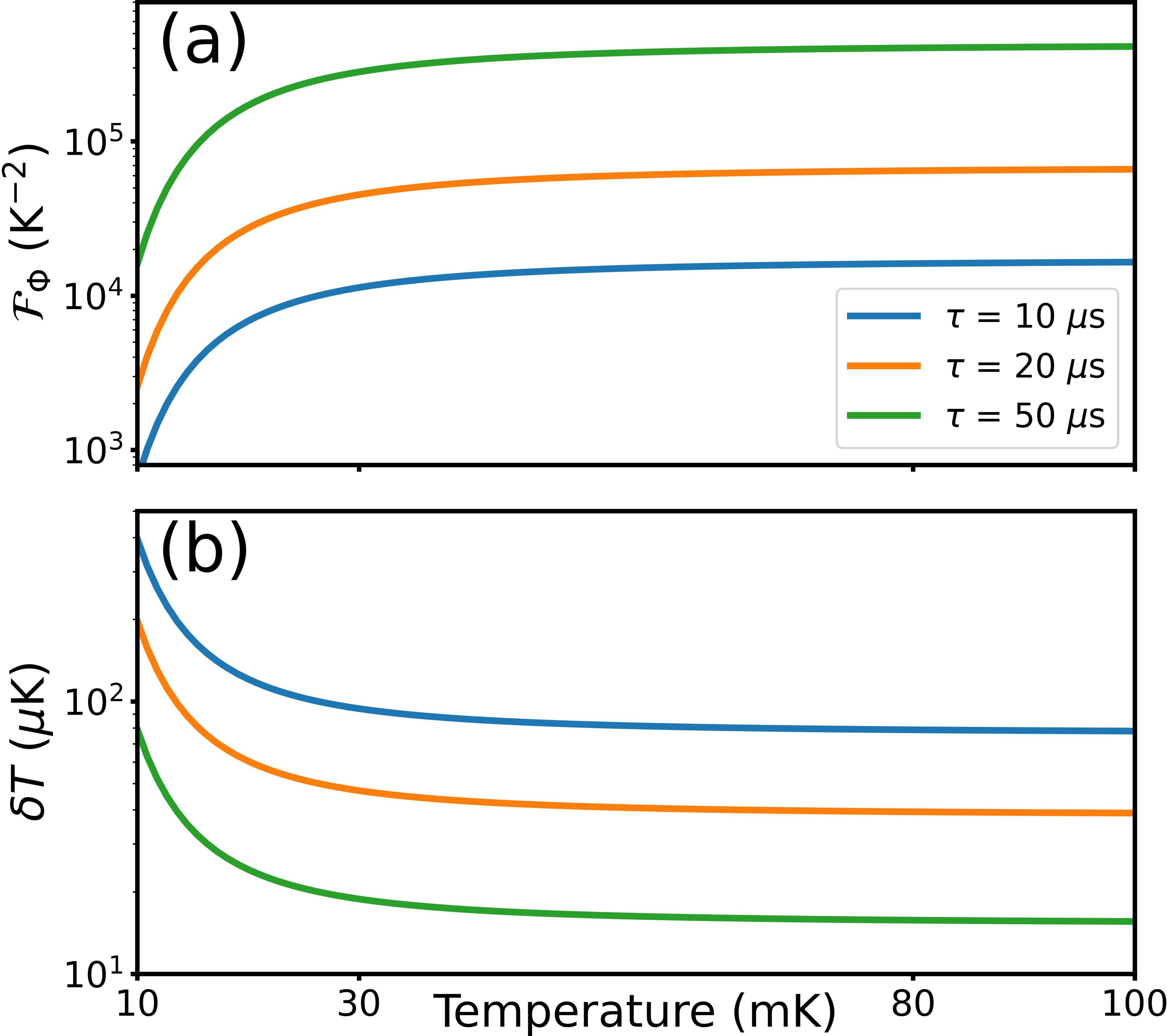} 
\caption{
(a) QFI $\mathcal{F}_\Phi(T)$ quantifies the temperature information encoded in the probe phase $\phi_b(T)=\lambda\tau\bar n_a(T)$, shown for interaction times $\tau=10,~20,$ and $50~\mu$s. 
At low temperatures ($k_BT\!\ll\!\hbar\omega_a$), $\mathcal{F}_\Phi$ is exponentially suppressed, while at high $T$ it saturates to a constant. Across curves, $\mathcal{F}_\Phi$ scales as $\propto\tau^2$. 
(b) Corresponding temperature resolution $\delta T(T)=1/\sqrt{\nu\,\mathcal{F}_\Phi(T)}$ for $\nu=10^4$ repetitions. 
Sensitivity improves as $1/\tau$ and flattens in the high-temperature limit.
}
\label{fig:p-qfi}
\end{figure}

Building on the qubit-based coherence measurement in Strategy 1, the same temperature-dependent phase $\phi_b(T)=\lambda\tau\bar n_a(T)$ can be viewed as the parameter that would accumulate during a Ramsey sequence on a dispersively coupled qubit.  
Alternatively, we measure the phase variation directly via heterodyne (or homodyne) detection of the high-$Q$ probe cavity, instead of mapping it onto a qubit superposition.

In this heterodyne-based implementation, the cross-Kerr interaction between the thermal mode and probe induces a temperature-dependent phase shift of the intracavity field.
This phase is faithfully transferred to the outgoing field and measured through heterodyne detection of its quadratures. 
Because the amplitude remains unchanged, the probe visibility is unity \(\mathcal{V}_\Phi(\tau,T)=1,\) and the information about temperature resides purely in the phase.  
The corresponding QFI,
\begin{equation}
    \mathcal{F}_\Phi(T) = 4 \alpha^2 \big(\lambda\tau\,\partial_T\bar n_a\big)^2,
\end{equation}
represents the ultimate precision bound for phase-based thermometry (Appendix~\ref{appendix:phase_qfi}).  
Figure~\ref{fig:p-qfi} shows $\mathcal{F}_\Phi(T)$ and the resulting temperature sensitivity $\delta T(T)=1/\sqrt{\nu\mathcal{F}_\Phi(T)}$ for several $\tau$.  
At low $T$, $\mathcal{F}_\Phi$ is exponentially suppressed, $\mathcal{F}_\Phi\!\sim\!T^{-4}e^{-2\hbar\omega_a/k_BT}$, while at high $T$ it saturates to $(\lambda\tau k_B/\hbar\omega_a)^2$.  
Longer interaction times simply rescale the vertical axis as $\mathcal{F}_\Phi\propto\tau^2$ and $\delta T\propto1/\tau$.

This approach leverages the exceptional phase stability of high-$Q$ 3D cavities ($\kappa_b\!\ll\!$ kHz) to permit millisecond-scale integration times without qubit-coherence limitations.  
Thus, it naturally supports continuous, low-bandwidth monitoring of slow thermodynamic drifts with state-of-the-art sensitivity.

Beyond steady-state thermometry, the same phase-tracking technique enables quantum calorimetry~\cite{karimi2020reaching}.  
A discrete energy deposition in the absorber produces a sudden jump $\Delta n_a$ in the thermal occupation, resulting in an instantaneous probe-phase step $\Delta\phi_b=\lambda\tau\Delta n_a$.  
Continuous heterodyne monitoring allows such steps to be time-tagged and resolved above the noise floor, enabling detection of rare, quantized energy arrivals.  
This connects our architecture to calorimetric applications in nuclear and high-energy physics~\cite{di2024quantum,doser2022quantum}, where resolving single-particle energy deposits with minimal backaction is essential.  
The long averaging time $\tau$ that optimizes steady-state sensitivity is, in the calorimetric mode, replaced by an effective integration window set by the probe bandwidth and digital filter, which together determine both energy resolution and timing jitter.

We therefore view the PS strategy as complementary to the CM approach.  
While Strategy 1 enables fast, high-bandwidth thermometry with short interaction times, Strategy 2 exploits long-lived cavity coherence for ultra-high sensitivity and naturally extends to calorimetry, where discrete energy events are registered as probe phase jumps.

\section{Experimental Implementation and Feasibility}\label{sec:experiment}

Building on the theoretical framework outlined above, we now describe a realistic experimental implementation of the CM quantum thermometry scheme. 
In addition to outlining a concrete device architecture, this section summarizes the main practical constraints at a coarse-grained level and how they enter the protocol. Importantly, these effects primarily reduce visibility and set an optimal interaction/mapping time, but do not alter the central operating principle of the scheme: a long-lived coherent reference mediates the readout of weak thermal fluctuations while the qubit remains largely protected during the sensing interval.

\subsection{Circuit Architecture and Mode Realization}\label{subsec:architecture}

The proposed thermometer comprises three essential elements: a high-coherence superconducting qubit, a low-frequency thermal mode, and a high-frequency coherent probe mode. 
These are integrated in a hybrid architecture combining planar CPW resonators with high-$Q$ 3D cavity components.
  
The probe is implemented as a 3D superconducting cavity in the $6{-}10$~GHz band with photon lifetimes in milli-second range~\cite{reagor2013reaching, romanenko2020three, kim2025ultracoherent}. 
Such longevity enables interaction times far exceeding qubit coherence times, thereby enhancing thermometric sensitivity. 
The long-lived probe state reduces the need for frequent reinitialization and allows repeated qubit interactions with minimal backaction. 
Compared to planar resonators, 3D cavities exhibit negligible internal dissipation, suppressing excess noise. 
Their dispersive coupling strength $\chi_b$ can be engineered geometrically (via qubit placement in the field) or spectrally (via detuning). 
They are also compatible with high-fidelity microwave readout, supporting both direct heterodyne detection and qubit-mediated measurements.
 
The thermal channel is realized as a planar CPW resonator or lumped LC circuit at $1{-}2$~GHz~\cite{zhu2022high, crowley2023disentangling}. 
Operating at lower frequencies ensures a sizable thermal population $\bar{n}_a$ even at tens of millikelvin. 
By coupling to an engineered bath, such as a terminated transmission line or resistor load, the mode rapidly equilibrates with its environment~\cite{gasparinetti2015fast, yeh2017microwave}, ensuring that it faithful tracks the bath temperature.

A superconducting transmon with typical coherence times of $T_1 \sim 300~\mu$s and $T_2^* \sim 100~\mu$s~\cite{place2021new, bal2024systematic, zhu2025disentangling} couples to the probe, allowing high-contrast Ramsey and Hahn-echo detection of probe dephasing and temperature mapping. 
Minimizing the coupling to the lossy thermal element remains essential to protect the qubit coherence.

In the CM scheme, the qubit remains in its ground state while thermal fluctuations imprint dephasing noise onto the long-lived probe during an interaction interval $\tau$. 
A subsequent Ramsey-type sequence maps this probe dephasing back to the qubit for readout (Appendix~\ref{app:mapping}). 
In the PS scheme, the probe field itself is monitored continuously via heterodyne detection; its ultra-high $Q$ enhances sensitivity to stochastic fluctuations in the thermal mode occupancy $\bar{n}_a(T)$.
In practice, heterodyne detection efficiency $\eta < 1$ reduces the observed Fisher information compared to the quantum limit, while homodyne measurement of a single quadrature can approach unity.

\subsection{Feasibility: Cross-Kerr Coupling and Practical Constraints}

In practice, feasibility is set by the need to generate measurable probe responses within realistic interaction times, while staying compatible with coherence times, cavity linewidths, and achievable qubit–cavity detunings in state-of-the-art superconducting platforms. 
These constraints define the operating window where both cross–Kerr couplings and dispersive pulls can be harnessed for thermometry in a competitive regime.

\subsubsection{Engineering Enhanced Cross--Kerr}\label{subsubsec:cross-Kerr}

\begin{figure}[b!]
\includegraphics[width=0.9\columnwidth]{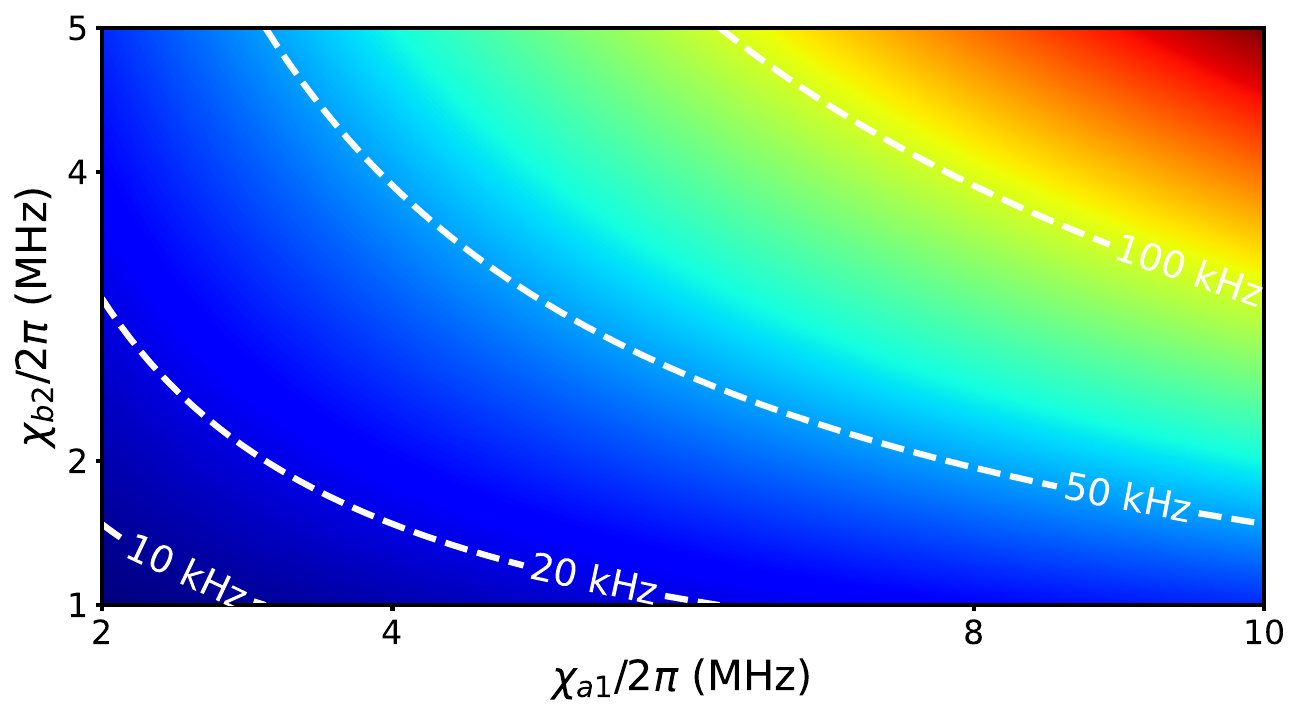} 
\caption{
Heatmap of the two-qubit--mediated cross--Kerr rate $\lambda/2\pi$ (kHz) versus the dispersive pulls $\chi_{a1}/2\pi$ and $\chi_{b2}/2\pi$ (MHz). 
Values are computed from $\lambda = 8\,\chi_{a1}\chi_{b2}\rho^2 / \Delta_{12}$ with $\rho = 0.2$ and $\Delta_{12}/2\pi=100~\mathrm{MHz}$. 
White dashed contours indicate $\lambda/2\pi=\{10,20,50,100\}$~kHz. 
}\label{fig:lambda_heatmap}  
\end{figure}

In the scheme, the effective cross–Kerr $\lambda$ can be engineered to imprint temperature–dependent fluctuations onto the probe within a realistic interaction time $\tau$, while remaining compatible with achievable detunings and coherence in a hybrid cQED architecture.

A single transmon dispersively coupled to both modes can generate a fourth-order virtual cross-Kerr interaction~\cite{blais2021circuit,zhang2019engineering}.
For realistic couplings and detunings this effective rate remains in the sub-kilohertz range, far too small for our sensing requirements, while attempts to enhance it by increasing hybridization inevitably degrade the probe’s quality factor and introduce additional Purcell and AC-Stark channels.

To overcome this limitation, we employ a passive two-qubit bridge (Q1, Q2) that links the thermal mode $a$ and the probe mode $b$ (Fig.~\ref{fig:schematic}).  
Mode $a$ couples dispersively to Q1 with pull $\chi_{a1}$; mode $b$ couples to Q2 with pull $\chi_{b2}$; and the qubits are capacitively linked through an exchange interaction $J_{XY}$. 
In the dispersive regime, \(|g_{a1,b2}| \ll |\Delta_{a1,b2}|, \qquad |J_{XY}| \ll |\Delta_{12}|,\) virtual transitions through the chain $a \!\leftrightarrow\! Q_1 \!\leftrightarrow\! Q_2 \!\leftrightarrow\! b$ generate an effective cross-Kerr (Appendix~\ref{app:cross_kerr})
\begin{equation}\label{eq:lambda_bridge_scaling}
\lambda = \frac{8 \chi_{a1}\chi_{b2}J_{XY}^2}{\Delta_{12}^3}.
\end{equation}
Here, we ignore the subleading terms in Eq.~\eqref{eq: cross-Kerr} that account for finite anharmonicity and detuning sign~\cite{blais2021circuit,didier2015fast}.  
Introducing the dimensionless ratio $\rho = J_{XY}/\Delta_{12}$, the scaling reduces to \(\lambda = 8\,\chi_{a1}\chi_{b2}\rho^2 / \Delta_{12}.\)
Figure~\ref{fig:lambda_heatmap} shows that a moderate $\lambda$ in $10$--$50$~kHz regime is easily reached by increasing $\chi_{a1}$ while maintaining modest $\chi_{b2}$ to protect the probe's quality factor.

A central advantage of the passive two-qubit bridge is that it leaves the 3D cavity nearly unaffected.  
With the state-of-the-art transmon lifetime \(T_1 \ge 300~\mu s\)~\cite{bal2024systematic, place2021new, zhu2025disentangling} and in the dispersive regime, the dominant added loss arises from Q2’s inverse-Purcell contribution, \(\kappa_{b,\mathrm{add}}/2\pi \approx (g_{b2}/\Delta_{b2})^2 / (2\pi T_1) \ll 1~\mathrm{kHz}\), far below the intrinsic linewidth of a high-$Q$ cavity.  
Unwanted $a\!\leftrightarrow\! b$ exchange is suppressed by ordering the qubit frequencies ($\omega_1<\omega_2$), leading to destructive interference between opposite-sign virtual paths.  
Meanwhile, the desired $\lambda$ remains large because it scales as $J_{XY}^2$.

For completeness, we note that alternative routes, such as parametric dressing, driven couplers, or flux-tunable bridges, also underscore the landscape of possible interaction-engineering tools to enhance cross-Kerr strength~\cite{blais2021circuit,gao2018programmable,greco2021quantum,zhou2023realizing,liu2020optimizing}. 
In general, these approaches introduce additional control complexity, Kerr drifts, or flux-sensitive dissipation, all of which need carefully engineered to reconcile with a high-$Q$ 3D cavity. 
Therefore, the passive two-qubit bridge may offer a stable, calibration-light, and hybrid cQED-compatible path to achieving strong cross-Kerr without compromising cavity coherence or introducing additional noise channels.

\subsubsection{Impact of Qubit-Mode Couplings on Visibility}

\begin{figure}[b!]
\centering
\includegraphics[width=0.9\columnwidth]{visibility.pdf}
\caption{
Qubit coherence visibility $\mathcal{V}_{\mathrm{R}}(\tau_R)$ as a function of the Ramsey mapping interval $\tau_R$ for different dispersive coupling strengths.
(a) Dependence on the probe-mode dispersive coupling $\chi_b/2\pi = 10,\,20,\,50~\mathrm{kHz}$. 
Larger $\chi_b$ accelerates photon-number-induced dephasing during the mapping and therefore shortens the usable Ramsey mapping window (solid curves). 
A Hahn-echo sequence (dashed curves) suppresses quasi-static photon-number fluctuations and significantly extends the allowable $\tau_R$.
(b) Dependence on the thermal-mode dispersive coupling $\chi_a/2\pi = 1,\,5,\,10~\mathrm{kHz}$. Maintaining small $\chi_a$ ensures that absorber fluctuations do not appreciably shorten the qubit Ramsey window.
}
\label{fig:visibility}
\end{figure}

During coherence mapping in the CM scheme, the qubit remains in its ground state throughout the interaction interval $\tau$, and is only promoted to a superposition during the short Ramsey (or Hahn-echo) window $\tau_R$ to sample the probe phase.  
Dephasing in this window originates from photon-number fluctuations of any dispersively coupled mode $m\in\{a,b\}$, with variance $\mathrm{Var}(n_m)$ and linewidth $\kappa_m$ (Appendix~\ref{appendix:parasitic}).  
The qubit Ramsey coherence envelope reads~\cite{gambetta2006qubit}
\begin{equation}
C_q =\exp\left\{-4\chi_m^2\mathrm{Var}(n_m)\left[\frac{\kappa_m\tau_R-1+e^{-\kappa_m\tau_R}}{\kappa_m^2}\right]\right\},
\label{eq:Cq}
\end{equation}
where $\mathrm{Var}(n_b)=|\alpha|^2+\bar n_b(1+\bar n_b)$ for the coherent probe and $\mathrm{Var}(n_a)=\bar n_a(1+\bar n_a)$ for the thermal absorber.  

Fig.~\ref{fig:visibility}(a) shows that stronger qubit-probe coupling $\chi_b$ increases the qubit dephasing rate for a modest probe population \(|\alpha|^2 = 1\), yielding a preserved mapping window $\tau_R \leq 1~\mu\mathrm{s}$ for the Ramsey envelope (solid lines). 
When a Hahn-echo sequence is applied, the corresponding filter function, \(F_{\mathrm{echo}}(\omega) = 8\sin^{4}(\omega \tau_R / 4) / \omega^{2},\) suppresses the low-frequency photon-number noise that dominates in high-$Q$ probe cavities ($\kappa_b \tau_R \ll 1$)~\cite{ithier2005decoherence,cywinski2008enhance,farfurnik2020characterizing}.  
Therefore, the echo reduces the effective dephasing rate by a factor $(\kappa_b \tau_R)^2/32$, restoring near-unity visibility even for $\chi_b/2\pi = 50~\mathrm{kHz}$ and $\tau_R$ can be up to $\sim 6~\mu\mathrm{s}$ (dashed lines).
Thus, fast mapping can be performed without appreciable loss of contrast, provided that the probe number fluctuations remain slow compared to $\tau_R$.

Fig.~\ref{fig:visibility}(b) illustrates that finite qubit-thermal coupling $\chi_a$ degrades visibility if present during mapping.  
Maintaining $\chi_a/2\pi\!\lesssim\!1$~kHz ensures that absorber fluctuations do not appreciably shorten the Ramsey window.

By contrast, the same echo operation in the qubit-only scheme would cancel the low-frequency absorber noise that encodes the temperature information, thereby recovering contrast but suppressing thermometric sensitivity.  

\subsection{Practical considerations and calibration}
We next outline general experimental considerations and practical operating procedures that can be used in future implementations to bound the achievable sensitivity and guide calibration.
In practice, achievable sensitivity is bounded by visibility loss during the mapping window, incomplete absorber thermalization, excess fluctuations, and systematic parameter errors. The mapping must be completed within the relevant coherence/linewidth scales, which yields a known visibility envelope and an optimal interaction time that maximizes the temperature-dependent slope (Eq.~\eqref{eq:Cq} and Fig.~\ref{fig:visibility}). Finite absorber equilibration to the engineered bath reduces the available temperature-dependent moments and can be captured by partially thermalized moments set by the absorber-bath coupling and the cycle time (Sec.~\ref{subsec:architecture}). Excess noise enters through the same absorber moments and can be modeled as additive offsets,
$\bar n_{a_1}(T)\!\rightarrow\!\bar n_{a_1}^{\rm th}(T)+\bar n_{\rm ex}$ and
$\mathrm{Var}(n_{a_1})\!\rightarrow\!\mathrm{Var}_{\rm th}(T)+\mathrm{Var}_{\rm ex}$,
calibrated via mapping-disabled reference runs (e.g., $\alpha\!=\!0$) and/or fits to known temperature points. Systematic errors (e.g., uncertainty in $\phi=\lambda\tau$ and residual higher-level/weak-drive effects) are mitigated by operating in the dispersive, weak-drive regime and by in situ calibrations of $\phi$ and mode populations (Sec.~\ref{subsubsec:cross-Kerr} and App.~\ref{app:cross_kerr}/\ref{appendix:parasitic}). These considerations do not alter the operating principle, but they set the optimal working point and provide a straightforward route to separating thermal signal from calibrated offsets.

\section{Comparative Performance of Quantum Thermometry Strategies}\label{sec:comparison}

\begin{figure}[b!]
\centering
\includegraphics[width=0.95\columnwidth]{Comparison.pdf}
\caption{
Comparison of quantum thermometry strategies at fixed cross-Kerr coupling $\lambda/2\pi = 50~\mathrm{kHz}$. 
(a) QFI $\mathcal{F}$ versus sensing time $\tau$ for the CM (blue), PS (orange), and QB (green) schemes. 
(b) Corresponding coherence envelope visibilities envelope $\mathcal{V}_{\mathrm{coh}}$: probe coherence for CM (blue), constant visibility for PS (orange),  and qubit Ramsey visibility for QB (green). 
Markers indicate the corresponding $\mathcal{V}_{\mathrm{coh}}$ at optimal $\tau^\star$ that maximizes $\mathcal{F}$ in CM and QB schemes.}
\label{fig:QFI_visibility_lambda}
\end{figure}

To benchmark performance, we compare the QFI $\mathcal{F}$ and coherence $\mathcal{V}_{\mathrm{coh}}$ across the three schemes as a function of the interaction time $\tau$.
Details of the QFI and coherence derivation of the qubit-only (QB) thermometry is presented in Appendix.~\ref{app:qubit_only_envelope}.
Figure~\ref{fig:QFI_visibility_lambda} summarizes the results for a fixed cross-Kerr coupling $\lambda/2\pi = 50~\mathrm{kHz}$ in the quasi-static regime $\kappa_a \tau \ll 1$.
This representation isolates the role of coherence decay and highlights the distinct mechanisms that determine each scheme’s optimal operating point.

The QB strategy (green) exhibits the characteristic rise-and-fall behavior expected when the qubit serves as both sensor and memory.  
For short interaction times, the QFI grows as $\mathcal{F} \propto \tau^2$ due to increased phase accumulation.  
At longer times, however, the same thermal fluctuations that encode temperature also induce qubit dephasing, suppressing the Ramsey visibility and causing $\mathcal{F}$ to collapse.  
Because this dephasing carries the temperature information itself, echo protocols cannot mitigate it without erasing the signal.  
The resulting optimal time $\tau^\star$ (green marker) defines a narrow high-information window.

The CM strategy (blue) exhibits a qualitatively similar peak-and-rolloff with $\tau$, but with a fundamentally different origin.  
Here the probe cavity, not the qubit, experiences temperature-induced phase diffusion.  
During the sensing interval the qubit is fully decoupled, therefore its intrinsic coherence is protected, allowing the probe to absorb all visibility loss.  
As $\tau$ increases, thermal fluctuations deepen the probe’s phase modulation, boosting $\mathcal{F}$ until probe dephasing dominates and visibility drops. 
The optimal $\tau^\star$ (blue marker) is therefore set by the probe quality factor and operating-point amplitude, both of which are experimentally tunable and ultimately constrained by measurement-induced backaction and finite readout SNR/contrast.
This separation between sensing probe and readout qubit enables CM to maintain high per-shot QFI without sacrificing qubit contrast.

The PS tracking scheme (orange) grows monotonically with $\tau$ at the rate expected from deterministic dispersive phase accumulation.  
Because the measurement does not rely on coherence of either the qubit or the probe, visibility remains essentially unity, and $\mathcal{F}_T$ lacks the peak-collapse structure associated with the coherence-based approaches.  
While its instantaneous QFI is lower at short times, the PS scheme can integrate indefinitely and is therefore limited primarily by measurement efficiency and low-frequency technical drifts.

For CM and QB schemes, the scaling $\mathcal{F}_T(\tau) \sim \tau^2 e^{-2\Gamma \tau}$ dictates an optimal interaction time \(\tau^\star \sim 1 / 2\Gamma\), where $\Gamma$ is the relevant dephasing rate (probe dephasing for CM, or qubit dephasing for QB).  
This defines an effective bandwidth $B^\star \sim \Gamma$ for rapid temperature tracking.  
By contrast, the PS strategy has $\mathcal{F}_T(\tau) \sim \tau^2$ without a visibility penalty, enabling long integrations but without the coherence-assisted gain available to CM.

The CM thermometry combines the short-time sensitivity of coherence-based sensing with resilience to the self-dephasing bottleneck that limits QB.  
The probe absorbs the visibility cost, the qubit retains full coherence contrast, and the QFI remains high across a broad temporal window.  
QB provides comparable sensitivity at longer times subsequently experiences rapid visibility collapse.  
PS is the most robust for long integrations but lacks the coherence-amplification available to CM.  
These results collectively position CM as the most versatile operating mode for quantum-limited thermometry, offering strong per-shot information and robustness to over-coupling and parameter drift.

\section{Outlook and Conclusion}\label{sec:summary}
The principle demonstrated here, using a coherent mode as an information buffer between a noisy environment and a fragile qubit, provides a key advantage by protecting the qubit's coherence while preserving temperature information in the buffered mode, thereby defining a general paradigm for coherence-mediated thermometry and noise spectroscopy. 
The regime of advantage for CMQT is the ultra-low-occupation limit $\bar n_a\!\ll\!1$, where conventional dispersive thermometry is fundamentally signal-limited; at higher temperatures the marginal benefit diminishes and simpler approaches can suffice.

In practice, the approach is fully compatible with state-of-the-art cQED hardware. 
The high-$Q$ 3D cavity can support millisecond coherence times, and the moderate cross-Kerr coupling can be engineered via multi-qubit pathways.
Beyond temperature estimation, closely related strategies can be extended to quantum calorimetry, fluctuation spectroscopy, and the characterization of correlated noise in superconducting quantum processors.
A natural next step is to cast the protocol in a multiparameter estimation framework, treating $T$ together with nuisance parameters such as effective absorber coupling rates, residual dispersive pulls, or detection efficiency; this would enable explicit robustness bounds and optimal experimental design under realistic calibration uncertainty.
More broadly, coherence-mediated sensing architectures offer a promising route for ultrasensitive calorimetry in nuclear and high-energy physics experiments, where millikelvin-scale energy resolution is increasingly critical.
Taken together, these considerations highlight both the broad applicability of coherence-mediated sensing and the immediate impact of our results for quantum thermometry.

In conclusion, we have introduced and analyzed a coherence-mediated thermometry scheme within a hybrid cQED architecture, in which thermal fluctuations are monitored indirectly through its cross-Kerr interaction with a coherent probe and read out via a dispersively coupled qubit.
By cleanly separating information acquisition from readout, the scheme circumvents the intrinsic self-dephasing bottleneck of qubit-based thermometers, enabling longer interaction times, higher per-shot sensitivity, and substantially improved robustness.
These results establish coherence-mediated quantum thermometry as a fundamentally distinct and scalable strategy, opening a path toward nondissipative, reusable quantum metrology with broad applicability across quantum science, precision sensing, and low-temperature detector technologies.

\noindent \textit{The data that support the findings of this article are openly available~\cite{python_script}.}

\begin{acknowledgments}
This work was supported by the U.S. Department of Energy, Office of Science, National Quantum Information Science Research Centers, Superconducting Quantum Materials and Systems Center (SQMS), under Contract No. 89243024CSC000002.
Fermilab is operated by Fermi Forward Discovery Group, LLC under Contract No. 89243024CSC000002 with the U.S. Department of Energy, Office of Science, Office of High Energy Physics.
\end{acknowledgments}

\appendix

\section{QFI from a Coherence Envelope}\label{app:QFI_definition}

We consider a qubit probe initialized in the superposition state \(|+\rangle=(|0\rangle+|1\rangle)/\sqrt{2}\), which is maximally sensitive to phase interaction. 
After an interaction time~$\tau$, the qubit undergoes pure dephasing due to temperature-dependent fluctuations, leading to the mixed state
\begin{equation}
\rho(T) = \frac{1}{2}
\begin{pmatrix}
1 & C\,e^{i\Phi}\\
C\,e^{-i\Phi} & 1
\end{pmatrix},
\end{equation}
where the temperature-dependent $C$ and $\Phi$ denote, respectively, the coherence envelope (the visibility of Ramsey fringes) and the mean accumulated phase. 
Both quantities generally depend on the interaction time $\tau$; here we suppress this dependence for brevity.

Physically, $C$ captures the loss of coherence induced by temperature-dependent fluctuations (e.g.\ photon-number noise in a coupled mode), while $\Phi$ represents a deterministic phase shift that can also carry temperature information. 
In this sense, $C$ quantifies the amplitude of the Bloch vector and $\Phi$ its orientation in the equatorial plane of the Bloch sphere.

The corresponding Bloch vector is
\[\mathbf r(T) = \big(C\cos\Phi,\, C\sin\Phi,\, 0\big),
\quad r = |\mathbf r| = C.\]
For a general qubit state $\rho=\tfrac{1}{2}(\mathbb{I}+\mathbf r\cdot\boldsymbol\sigma)$, the quantum Fisher information (QFI) for estimating a parameter $T$ is known to be~\cite{liu2020quantum, zhong2013fisher}
\begin{equation}
\mathcal{F}_T 
= \frac{\big(\partial_T r\big)^2}{1-r^2} 
+ r^2\big(\partial_T \phi\big)^2,
\label{eq:QFI-qubit-master}
\end{equation}
with \(r = C, \mathrm{and} \phi=\Phi\). 
This expression separates the two distinct resources available for thermometry: changes in the coherence amplitude $C$ and changes in the phase angle $\Phi$. These are two complementary information channels in the likelihood: amplitude/visibility changes and phase shifts.
Explicitly,
\begin{equation}
\mathcal{F}_T 
= \frac{\big(\partial_T C\big)^2}{1-C^2} 
+ C^2\big(\partial_T \Phi\big)^2.
\label{eq:QFI-main}
\end{equation}
The first term arises because the eigenvalues of $\rho(T)$, $\lambda_\pm=\tfrac{1}{2}(1\pm C)$, depend on $T$ through $C(T)$.  
The second term originates from the $T$-dependence of the eigenvectors, which rotate with $\Phi(T)$. 
Thus, loss of coherence (shrinking Bloch vector) and coherent phase shifts (rotation) both provide temperature sensitivity, but they contribute in qualitatively different ways.

In particular, if the accumulated phase $\Phi$ is independent of $T$, the QFI reduces to a purely dephasing-based form:
\begin{equation}
\mathcal{F}_T = \dfrac{\big(\partial_T C\big)^2}{1-C^2}.
\label{eq:QFI-envelope-only}
\end{equation}
This regime describes thermometry strategies where temperature information is encoded solely in the coherence envelope of the qubit.

Although the above derivation was formulated for a qubit undergoing temperature-dependent dephasing, the resulting expression is fully general for any coherence channel characterized by a complex visibility $Ce^{i\Phi}$. 
In the main text, this formalism is applied to the probe mode: the absorber’s temperature modifies the probe’s complex amplitude in the same way, with $C$ describing the loss of visibility and $\Phi$ the temperature-dependent phase shift. 
The qubit in that scheme serves only as a phase-sensitive transducer that converts the probe’s coherence into a measurable Ramsey signal. 
Therefore, Eq.~(\ref{eq:QFI-main}) quantifies the fundamental temperature information accessible in both the qubit- and probe-based representations.

\section{Derivation of the Coherence-Mediated Envelope and QFI}\label{app:C_envelope}

To model the temperature-dependent coherence decay observed in our qubit, we derive the expression for the coherence envelope $C(\tau, T)$ starting from a microscopic picture of phase diffusion induced by thermal fluctuations acting on a coherent state.

We consider a coherent probe mode initially prepared in the state $|\alpha\rangle$, coupled via a cross-Kerr interaction to a thermal mode:
\begin{equation}
H_{\text{int}} = \lambda \hat{n}_a \hat{n}_b,
\end{equation}
where $\lambda$ is the cross-Kerr rate, and $\hat{n}_a$, $\hat{n}_b$ are photon-number operators for the thermal and coherent modes, respectively.  

During an interaction of duration $\tau$, the coherent state accumulates a phase shift depending on the instantaneous thermal photon number:
\[
|\alpha\rangle \;\rightarrow\; |\alpha e^{i\phi_b}\rangle, \quad \phi_b = \lambda n_a \tau.
\]
Because $n_a$ fluctuates in a thermal state, the coherent probe undergoes random phase shifts drawn from a distribution set by the thermal statistics.

For a thermal state, $n_a$ follows a Bose–Einstein distribution with mean $\bar{n}_a$ and variance $\mathrm{Var}(n_a) = \bar{n}_a (1+\bar{n}_a)$. The resulting phase variance is
\begin{equation}
\sigma_\phi^2 = \mathrm{Var}(\phi) = (\lambda \tau)^2 \bar{n}_a (1+\bar{n}_a).
\end{equation}
Thus, the probe mode experiences Gaussian phase diffusion, described by a phase-averaged state:
\begin{equation}
\rho_b(T) = \int \! d\phi\, P(\phi)\, |\alpha e^{i\phi}\rangle \langle \alpha e^{i\phi}|,
\end{equation}
with \(P(\phi) = \tfrac{1}{\sqrt{2\pi\sigma_\phi^2}}\, e^{-\phi^2/2\sigma_\phi^2}.\)

The probe coherence envelope is proportional to the overlap of the original probe state with the diffused mixture:
\begin{equation}
C(\tau, T) = \langle \alpha | \rho_b(T) | \alpha \rangle 
           = \int d\phi\, P(\phi)\, \big|\langle \alpha | \alpha e^{i\phi} \rangle\big|^2.
\end{equation}
We obtain
\begin{align*}
|\langle \alpha | \alpha e^{i \phi} \rangle|^2 
&= \exp\!\big[ -2|\alpha|^2(1-\cos\phi) \big] \\
&= e^{-2\alpha^2} e^{2\alpha^2\cos\phi}.
\end{align*}
using \(\langle \alpha | \beta \rangle = \exp\!\left[\, \alpha^* \beta - \tfrac{1}{2}\big(|\alpha|^2 + |\beta|^2\big)\right]\) and $\beta = \alpha e^{i\phi}$.

Substituting back, we have
\begin{equation}\label{eq:C-define}
C(\tau,T) = e^{-2\alpha^2}\int d\phi\, P(\phi)\, e^{2\alpha^2\cos\phi}.
\end{equation}
The integral in Eq.~\eqref{eq:C-define} can be evaluated in closed form using the Gaussian statistics of $\phi$:
\begin{equation}
C(\tau,T) = \exp\!\left[-2\alpha^2\left(1-e^{-\sigma_\phi^2/2}\right)\right].
\end{equation}
We define the effective dephasing strength as
\begin{equation}
\Gamma_\phi(T) = \sigma_\phi^2/2 = \frac{1}{2}(\lambda\tau)^2 \bar{n}_a (1+\bar{n}_a).
\end{equation}

The final coherence envelope is therefore
\begin{equation}\label{eq:cm_envelope}
C(\tau, T) = \exp\!\big[-2\alpha^2 (1-e^{-\Gamma_\phi})\big].
\end{equation}
At low temperatures ($\bar{n}_a\to 0$), the dephasing vanishes and $C(\tau,T)\to 1$. At high temperatures, the envelope saturates to $\exp(-2\alpha^2)$, corresponding to complete phase scrambling. 
The nonlinear dependence on both $\alpha$ and $\bar{n}_a$ makes the scheme sensitive to thermal fluctuations even deep in the sub-photon regime. 

In the CM protocol, all $T$-dependence arises from the envelope $C(\tau,T)$; there is no temperature-dependent phase offset $\Phi(T)$. 
Differentiating Eq.~\eqref{eq:cm_envelope} gives
\begin{equation}
\partial_T C
= C \cdot 2\alpha^2 e^{-\Gamma_\phi}\, \partial_T \Gamma_\phi,
\end{equation}
with
\(\partial_T \Gamma_\phi = \tfrac{1}{2}(\lambda\tau)^2 (1+2\bar{n}_a) \partial_T \bar{n}_a,\) and 
\(\partial_T \bar{n}_a = \hbar\omega_a\,\bar{n}_a (1+\bar{n}_a) / k_B T^2.
\)

Substituting in Eq.~\eqref{eq:QFI-envelope-only}, we obtain
\begin{equation}\label{eq: CM-QFI}
\mathcal{F}_T(\tau,T)
= \frac{\Big[\alpha^2 e^{-\Gamma_\phi} (\lambda\tau)^2 \,(1+2\bar{n}_a)\,\partial_T \bar{n}_a\Big]^2 \, C^2}{1-C^2}.
\end{equation}
In the low-occupation and weak-dephasing limit ($\bar{n}_a \ll 1$ and $\Gamma_\phi \ll 1$), Eq.~\eqref{eq: CM-QFI} simplifies to
\begin{equation}
\mathcal{F}_T(\tau,T) \;\approx\; \alpha^2 (\lambda\tau)^2 \left(\frac{\hbar\omega_a}{k_B T^2}\right)^2 \bar{n}_a.
\end{equation}
This shows that the QFI scales linearly with thermal occupation $\bar{n}_a(T)$, quadratically with the effective interaction strength $\lambda\tau$, and linearly with the probe photon number $|\alpha|^2$, highlighting the bosonic amplification inherent in the CM approach.

\section{Derivation of Phase-Shift QFI}\label{appendix:phase_qfi}
 
During the interaction time~$\tau$, thermal fluctuations in mode~$a$ induce a temperature-dependent frequency (and hence phase) shift on the probe mode~$b$ via the cross-Kerr coupling~$\lambda$. 
The outgoing probe field, monitored by a heterodyne detector, acquires a mean phase
\begin{equation}
    \phi_b(T) = \lambda \tau\, \bar{n}_a(T),
\end{equation}
where $\bar{n}_a(T)$ is the thermal occupation of the absorber mode.  
The measurement outcome is the complex probe quadrature $b_{\text{out}} = |\alpha| e^{i\phi_b(T)}$, whose phase carries the temperature information.

The temperature-dependent phase shift transforms an initial coherent probe $|\alpha\rangle$ into \(|\psi(T)\rangle = |\alpha e^{i\phi_b(T)}\rangle .\)
The quantum Fisher information (QFI) for temperature estimation from this pure state is~\cite{paris2009quantum, braunstein1994statistical, toth2014quantum}
\begin{equation}
    \mathcal{F}_\Phi(T)
    = 4\!\left[ \langle \partial_T \psi | \partial_T \psi \rangle
    - \big| \langle \psi | \partial_T \psi \rangle \big|^2 \right].
\end{equation}
Using $\partial_T|\psi\rangle = i(\partial_T\phi_b)\,\hat{n}_b|\psi\rangle$ and the photon-number variance 
$\mathrm{Var}(\hat{n}_b)=|\alpha|^2$ for a coherent state, we find
\begin{equation}\label{eq:F_phi_result}
    \mathcal{F}_\Phi(T) = 4|\alpha|^2 \big( \partial_T \phi_b(T) \big)^2 = 4|\alpha|^2 \big( \lambda \tau \, \partial_T \bar{n}_a \big)^2.
\end{equation}
     
Equation~\eqref{eq:F_phi_result} shows that the PS strategy benefits from a simple, fully coherent mapping: the temperature dependence enters only through the mean phase $\phi_b(T)$, yielding a clean $|\alpha|^2\tau^2$ scaling that can be boosted by increasing probe power or interaction time.  
At the same time, the sensitivity diminishes at low temperature through the factor $\partial_T\bar n_a$, reflecting the reduced thermal susceptibility of the absorber mode.  
Thus, this scheme offers a straightforward and hardware–efficient thermometric readout, albeit with a narrower useful window than coherence-based sensing.

\section{Derivation of Qubit-Only-Based Coherence Envelope and QFI}\label{app:qubit_only_envelope}

\begin{figure}[b!]
\includegraphics[width=0.95\columnwidth]{QB-QFI.pdf} 
\caption{
(a) QFI $\mathcal{F}_q(T,\tau)$ as a function of temperature $T$ and interaction time $\tau$ for a representative dispersive coupling $\chi_a/2\pi = 50~\mathrm{kHz}$. 
The maximum $\mathcal{F}_q$ and corresponding optimal $\tau^\star$ both decrease as the thermal occupation increases (white dashed line).
(b) $\mathcal{F}_q$ versus $\tau$ for fixed absorber linewidths $\kappa_a/2\pi = 10,\,10^2,\,10^3~\mathrm{kHz}$ at $T = 10~\mathrm{mK}$. 
Small $\kappa_a$ (quasi-static limit) yields a finite optimal $\tau^\star$, while large $\kappa_a$ produces Markovian dynamics for which $\mathcal{F}_q(\tau)$ grows monotonically.
(c) $\mathcal{F}_q$ versus $\tau$ for different dispersive couplings $\chi_a/2\pi = 10,\,100~\mathrm{kHz}$. 
Increasing $\chi_a$ compresses the optimal sensing window to shorter $\tau^\star$, while the peak value of $\mathcal{F}_q$ remains nearly unchanged, indicating that stronger coupling mainly reshapes the per-shot information rather than enhances it.}
\label{fig:q_qfi}
\end{figure}

We consider a qubit dispersively coupled to a single harmonic mode $\hat a$ with frequency $\omega_a$ and linewidth $\kappa_a$.
To maintain consistency with the definition used in the CM scheme, we write the dispersive interaction in longitudinal form
\[
    \hat H_{\mathrm{int}} = \tfrac{1}{2}\,\chi_a\, \hat a^\dagger \hat a \,\hat\sigma_z,
\]
so that $\chi_a$ denotes the qubit frequency shift per thermal photon.

For thermometry, the relevant effect of the thermal mode is the stochastic modulation of its photon number.
We treat $\delta\hat n_a(t)=\hat a^\dagger(t)\hat a(t)-\bar n_a$ as a stationary random process fully characterized by its two-point correlator.
In the weak-dispersive and weak-dephasing regime, the qubit's phase evolution is accurately captured by the second-order cumulant (Gaussian-phase) approximation \cite{clerk2010introduction, gambetta2006qubit, blais2021circuit}.
We define the thermal correlation as 
\begin{equation}
    C_{nn}(\tau) = \langle \delta\hat n_a(\tau)\,\delta\hat n_a(0)\rangle = \bar n_a(\bar n_a+1)\, e^{-\kappa_a|\tau|}.
\end{equation}

During a Ramsey evolution of duration $\tau$, the qubit accumulates the stochastic phase
\begin{equation}
    \phi(\tau) = \chi_a\!\int_0^\tau dt\, \hat n_a(t) = \chi_a\bar n_a\,\tau + \chi_a\!\int_0^\tau dt\,\delta\hat n_a(t).
\end{equation}
The cumulant expansion yields the coherence
\begin{equation}
    C_q(\tau) = \exp[-i\chi_a\bar n_a\,\tau]\,\exp\!\Big[-\tfrac12\,\langle\!\langle\phi_{\rm noise}^2(\tau)\rangle\!\rangle\Big],
\end{equation}
with
\[
    \langle\!\langle \phi_{\rm noise}^2(\tau)\rangle\!\rangle = \chi_a^2 \int_0^\tau\!\!\int_0^\tau dt_1\,dt_2\,C_{nn}(t_1-t_2).
\]
Carrying out the double integral yields
\[
    \int_0^\tau\!\!\int_0^\tau dt_1 dt_2\,e^{-\kappa_a|t_1-t_2|} = \frac{2(\kappa_a\tau - 1 + e^{-\kappa_a\tau})}{\kappa_a^2}.
\]
The Ramsey coherence envelope is therefore
\begin{equation}\label{eq:C_qb}
    C_q(\tau) = \exp[-i\chi_a\bar n_a\,\tau]\,\exp\!\left[-\chi_a^2\,\bar n_a(\bar n_a+1)\, f_\kappa(\tau)\right],
\end{equation}
with \(f_\kappa(\tau)=(\kappa_a\tau - 1 + e^{-\kappa_a\tau}) / \kappa_a^2.\)
This expression interpolates between the quasi-static regime $\kappa_a\tau\ll 1$, where $f_\kappa(\tau)\approx \tfrac12\tau^2$, and the Markovian regime $\kappa_a\tau\gg1$, where $f_\kappa(\tau)\approx \tau/\kappa_a$.

From Eq.~\eqref{eq:C_qb}, the coherence amplitude is $|C_q(\tau,T)| = e^{-\Gamma_q}$, with \(\Gamma_q = \chi_a^2\,\bar n_a(\bar n_a+1)\, f_\kappa(\tau),\) and the accumulated phase is $\Phi_q(\tau,T)=\chi_a\bar n_a\,\tau$.  
Differentiation yields \(\partial_T \Gamma_q = \chi_a^2 f_\kappa(\tau)\,(1+2\bar n_a)\,\partial_T\bar n_a\) and \(\partial_T \Phi_q = \chi_a \tau\,\partial_T \bar n.\)

Substituting these into the pure-state QFI expression gives
\begin{equation}
\mathcal{F}_q = \frac{(\partial_T \Gamma_q)^2}{e^{2\Gamma_q}-1}\;+\; e^{-2\Gamma_q}\,\big(\chi_a \tau\big)^2\,(\partial_T\bar n_a)^2.
\end{equation}

Figure~\ref{fig:q_qfi} illustrates the resulting behavior. 
For a representative dispersive coupling $\chi_a / 2\pi = 50~\mathrm{kHz}$, the qubit-only QFI $\mathcal{F}_q$ exhibits the same qualitative trends as the CM scheme in the quasi-static limit ($\kappa_a \tau \ll 1$): the optimal $\mathcal{F}_q$ and corresponding interaction time $\tau^\star$ both decrease with increasing bath temperature $T$, as shown in Fig.~\ref{fig:q_qfi}(a).
Fig.~\ref{fig:q_qfi}(b) shows that in the quasi-static regime (small $\kappa_a$) the competition between information gain and dephasing yields a finite optimal interaction time $\tau^\star$, whereas for large $\kappa_a$ the dynamics become effectively Markovian and $\mathcal{F}_q(\tau)$ increases monotonically with~$\tau$.
As shwon in Fig.~\ref{fig:q_qfi}(c), stronger dispersive couplings $\chi_a$ drive the optimum to shorter times $\tau^\star$, but the maximum achievable $\mathcal{F}_q$ remains nearly independent of $\chi_a$, indicating that stronger coupling primarily compresses the optimal sensing window rather than enhancing the overall per-shot information.

Importantly, in the QB scheme the qubit is both sensor and memory: the same dispersive coupling that encodes temperature also dephases the qubit.  
This intrinsic trade-off cannot be removed by echo without erasing the signal, in contrast to the coherence-mediated scheme where the temperature-dependent diffusion is absorbed entirely by the probe mode.
Thus, practical operation requires choosing $\tau\approx\tau^\star$ and balancing the gain from large $\chi_a$ against the inevitable loss of visibility. 
In contrast, the CM scheme off-loads the sensing burden onto a high-$Q$ probe mode: while the probe collects the temperature-dependent phase diffusion, the qubit can be dynamically decoupled from it, enabling much longer interaction times and leveraging the stability of a 3D cavity to preserve visibility.

\section{Sixth-order perturbation theory for the two--qubit mediated cross--Kerr}
\label{app:cross_kerr}

To make the virtual process transparent, we evaluate the cross-Kerr coupling between bosonic modes $a$ and $b$ using straightforward sixth--order perturbation theory (PT)~\cite{Griffiths2018}. Two transmon qubits $Q_1$ and $Q_2$ bridge the modes in series ($a$-$Q_1$-$Q_2$-$b$). 

The interaction term of the system Hamiltonian is 
\begin{align*}
V/\hbar = \,& g_{a1}(a\sigma_1^+ + a^\dagger\sigma_1^-) + g_{b2}(b\sigma_2^+ + b^\dagger\sigma_2^-)\; + \\
            & J_{XY}(\sigma_1^+\sigma_2^- + \sigma_1^-\sigma_2^+).
\end{align*}
In the dispersive regime, \(|\Delta_{a1, b2}| \gg |g_{a1, b2}|, \text{and} |\Delta_{12}| \gg |J_{XY}| \), with detunings \(\Delta_{a1} = \omega_a - \omega_1, \Delta_{b2} = \omega_b - \omega_2, \Delta_{12} = \omega_1 - \omega_2.\)

The connected 6th-order perturbation theory for the energy correction can be calculated as:
\begin{align*}
\Delta E_i^{(6)} &= \sum_{j,k,l,m,n\neq i}\frac{\bra{i}V\ket{n}\bra{n}V\ket{m}\bra{m}V\ket{l}}{(E_i-E_n)(E_i-E_m)(E_i-E_l)}\\[4pt]
                 &\qquad\times\frac{\bra{l}V\ket{k}\bra{k}V\ket{j}\bra{j}V\ket{i}}{(E_i-E_k)(E_i-E_j)}.
\end{align*}
where \(\ket{i} = \ket{1_a, 1_b, gg}\) is the initial state, and the sum is over all possible sequences of intermediate virtual states \(\ket{j, k, l, m, n}\).

To calculate the energy of the state \(\ket{1_a, 1_b, gg}\) (one photon in each mode and both qubits in ground state) and find the part of its energy that depends on both photons being present, we list one of possible virtual coupling paths:
\begin{enumerate}[itemsep=-1mm,topsep=0mm]
    \item \(\ket{1_a, 1_b, gg} \xrightarrow{g_{a1} a \sigma_1^+} \ket{0_a, 1_b, eg}\): a photon from mode $a$ virtually excites $Q_1$;
    \item \(\ket{0_a, 1_b, eg} \xrightarrow{g_{b2} b \sigma_2^+} \ket{0_a, 0_b, ee}\): a photon from mode $b$ virtually excites $Q_2$;
    \item \(\ket{0_a, 0_b, ee} \xrightarrow{g_{b2} b^\dagger \sigma_2^-} \ket{0_a, 1_b, eg}\): $Q_2$ de-excites, returning the photon to $b$;
    \item \(\ket{0_a, 1_b, eg} \xrightarrow{J_{XY} \sigma_1^- \sigma_2^+} \ket{0_a, 1_b, ge}\): flip the excitation from $Q_1$ to $Q_2$;
    \item \(\ket{0_a, 1_b, ge} \xrightarrow{J_{XY} \sigma_1^+ \sigma_2^-} \ket{0_a, 1_b, eg}\): flip the excitation from $Q_2$ to $Q_1$;
    \item \(\ket{0_a, 1_b, eg} \xrightarrow{g_{a1} a^\dagger \sigma_1^-} \ket{1_a, 1_b, gg}\): $Q_1$ de-excites, returning the photon to $a$.
    \vspace{2mm}
\end{enumerate}

Collecting these steps, the representative ladder can be summarized as:
\begin{align*}
&\ket{11,gg}
\xrightarrow{g_{a1}} \ket{01,eg}
\xrightarrow{g_{b2}} \ket{00,ee}
\xrightarrow{g_{b2}} \ket{01,eg}
\xrightarrow{J_{XY}} \\ &\ket{01,ge}
\xrightarrow{J_{XY}} \ket{01,eg}
\xrightarrow{g_{a1}} \ket{11,gg}.
\end{align*}
The product of the matrix elements (numerator) for the path is straightforward
\begin{align*}
 &\bra{i}V\ket{n} \bra{n}V\ket{m} \bra{m}V\ket{l} \bra{l}V\ket{k}  \bra{k}V\ket{j} \bra{j}V\ket{i}\\
 &= g_{a1}\, J_{XY} \, J_{XY} \, g_{b2}\, g_{b2}\, g_{a1} \\
 &= g_{a1}^2\, g_{b2}^2\, J_{XY}^2.   
\end{align*}
Each intermediate state contributes an energy denominator
$E_i-E_{j, k, l, m, n}$:
\begin{align*}
&\ket{01,eg}: \Delta_{a1};\;
\ket{01,ge}: \Delta_{a1}+\Delta_{12};\;
\ket{00,ee}: \Delta_{a1}+\Delta_{b2}; \\
&\ket{01,ge}: \Delta_{a1}+\Delta_{12};\;
\ket{01,eg}: \Delta_{a1},    
\end{align*}
and the denominator product is
\[
\mathcal{D}=\Delta_{a1}^2\,(\Delta_{a1}+\Delta_{12})^2\,(\Delta_{a1}+\Delta_{b2}).
\]

The derivation above follows the ladder where mode $a$ excites $Q_1$ first. There exists a mirror ladder where mode $b$ excites $Q_2$ first, producing an analogous expression with indices $a_1 \leftrightarrow b_2$. Therefor, the 6th–order correction contributes to an effective cross–Kerr interaction
\[
\lambda\;\sim\;\frac{g_{a1}^2 g_{b2}^2 J_{XY}^2}{\Delta_{a1}^2 (\Delta_{a1}+\Delta_{12})^2 (\Delta_{a1}+\Delta_{b2})}.
\]
Additionally, there are other time orderings permitted by Rayleigh–Schrödinger perturbation theory. Summing all allowed ladders restores a symmetric expression in $(\Delta_{a1}, \Delta_{b2}, \Delta_{12})$, and gives the effective cross-Kerr
\begin{equation}\label{eq: cross-Kerr}
    \lambda = \frac{8 \chi_{a1} \chi_{b2} J_{XY}^2}{\Delta_{12}^3}\, \Big[1+\mathcal{O}\!\big(\tfrac{\Delta_{12}}{\Delta_{a1}},\;\tfrac{\Delta_{12}}{\Delta_{b2}}\big)\Big],
\end{equation}
with \(\chi_{a1} = - g_{a1}^2 / \Delta_{a1}\) and \(\chi_{b2} = - g_{b2}^2 / \Delta_{b2}\). Here, the subleading terms account for finite anharmonicity and detuning sign~\cite{blais2021circuit,didier2015fast}.

In practice, the same cross–Kerr interaction can be gated by flux-tuning one of the bridge qubits (or the intermediate coupler) to temporarily reduce the effective detuning $\Delta_{12}$ during the sensing window.  
Such a flux-tunable transmon implementation enables transiently enhanced $J_{XY}/\Delta_{12}$ while the system remains dispersive in its idle configuration, thereby enhancing the cross–Kerr rates without compromising qubit coherence.

It is noteworthy that the same cross-Kerr interaction that enables temperature sensing also mediates measurement backaction: fluctuations of the probe photon number $\hat{n}_b$ induce an additional dephasing rate in the absorber,
\(\Gamma_{\phi,a} \propto \lambda^2 \bar{n}_b / \kappa_b\),
which manifests as extra decay in the probe's coherence envelope during the interaction~\cite{rigetti2012superconducting}.  
Minimizing this backaction constrains the allowable probe photon number $\bar{n}_b = \alpha^2$ and defines the optimal balance between signal transduction and absorber disturbance.  
In a gated implementation, a flux--tunable bridge can be activated only during the sensing window, thereby limiting the accumulated backaction while preserving the desired transient cross-Kerr strength.

\section{Parasitic Dispersive Couplings and Qubit Dephasing}
\label{appendix:parasitic}

While the cross-Kerr interaction between the probe ($b$) and thermal ($a$) modes is the desired sensing channel, the qubit also retains residual dispersive couplings to each mode, with strengths $\chi_b$ and $\chi_a$. These couplings do not encode thermometric information; instead, they generate parasitic dephasing during the Ramsey readout window of duration~$\tau_R$.

If the qubit couples dispersively to a bosonic mode $m$ with photon-number variance $\mathrm{Var}(n_m)$ and linewidth $\kappa_m$, a second-order cumulant expansion gives the general coherence envelope during the Ramsey window~\cite{gambetta2006qubit}:
\begin{equation}
\label{eq:Cq_general}
C_q^{R}(\tau_R) = \exp\!\left[-\,4\chi_m^{\,2}\,\mathrm{Var}(n_m)\, f^R(\kappa_m, \tau_R) \right],
\end{equation}
with \(f^R(\kappa_m, \tau_R) = (\kappa_m \tau_R - 1 + e^{-\kappa_m \tau_R}) / \kappa_m^{2}.\)
This expression captures the full crossover between slowly fluctuating and rapidly fluctuating photon-number noise in a single closed form.

For the thermal mode, linewidth $\kappa_a$ and temperature-dependent variance $\mathrm{Var}(n_a)$ jointly determine the associated dephasing rate, but this pathway plays no constructive role in sensing and should be mitigated by engineering $\chi_a$ to be small.

For the probe mode, which is initialized in a displaced thermal state, the photon-number variance is
\(\mathrm{Var}(n_b)= \bar n_b (1+\bar n_b) + |\alpha|^2(1+2\bar n_b),\)
Increasing $|\alpha|$ enhances temperature-to-phase transduction in the probe but also increases parasitic qubit dephasing through the residual coupling $\chi_b$ during the short Ramsey readout. 

Therefore, applying a Hahn echo during the readout window yields a modified coherence envelope that suppresses quasi-static photon-number fluctuations in the probe. For dispersive coupling to mode $b$, the exact echo envelope is~\cite{farfurnik2020characterizing, ithier2005decoherence, cywinski2008enhance}
\begin{equation}
C_{q}^E(\tau_R) = \exp\!\left[-\,4\chi_b^{\,2}\,\mathrm{Var}(n_b)\, f^E(\kappa_b, \tau_R) \right],
\label{eq:Cq_echo}
\end{equation}
with \(f^E(\kappa_b, \tau_R) = (\kappa_b \tau_R - 3 + 4e^{ \kappa_b \tau_R/2} - e^{-\kappa_b \tau_R}) / \kappa_b^{2}.\)

All parasitic pathways enter through the unified coherence laws in Eqs.~\eqref{eq:Cq_general}-\eqref{eq:Cq_echo}, with distinct photon statistics for the probe and thermal modes.  
Engineering small $\chi_b$ and $\chi_a$ ensures that the qubit maintains high visibility during the brief Ramsey readout, while the probe amplitude $|\alpha|$ and the cross-Kerr interaction remain free parameters for optimizing thermometric sensitivity. 
Echo sequences can be used to suppress quasi-static probe-induced fluctuations arising from the high-$Q$ probe cavity.

\section{Phase-to-Quadrature Mapping}\label{app:mapping}

\begin{figure}[t]
  \centering
  \includegraphics[width=0.9\columnwidth]{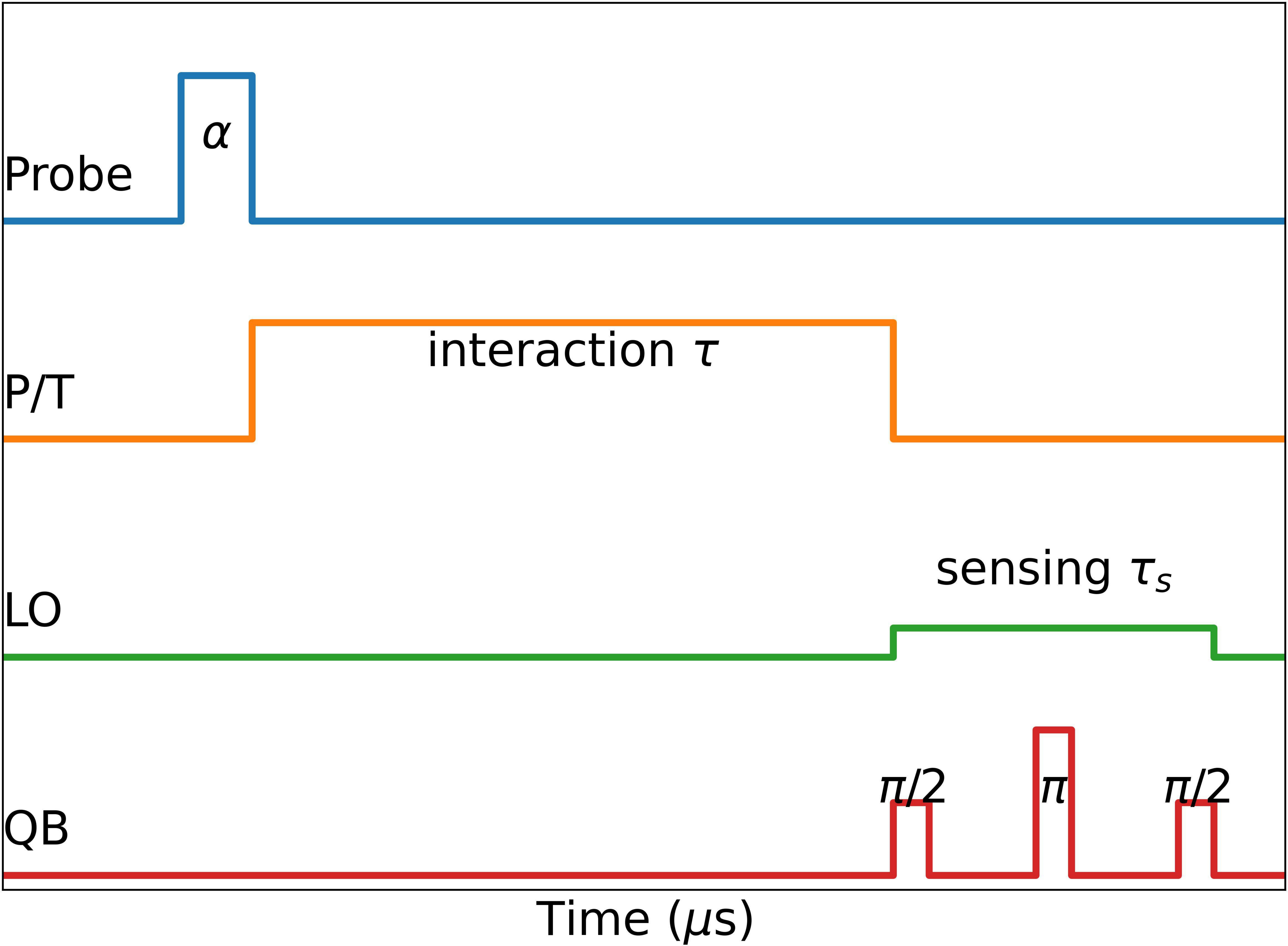}
  \caption{Phase–to–quadrature mapping pulse sequence. 
A short pulse prepares the high-$Q$ 3D cavity (probe) in a coherent state of amplitude $\alpha$ (blue). 
The probe then interacts with the thermal mode $a$ for a duration $\tau$, acquiring a phase 
$\phi_b=\lambda\!\int_0^\tau n_a\,dt$ (yellow), while the qubit remains idle. 
Next, a short sensing window of length $t_R$ turns on the qubit–probe dispersive coupling $\chi_b$ (green). 
During this window a weak LO displacement $\beta=|\beta|e^{i\theta}$ is applied to set the analyzed quadrature (green). 
A Hahn-echo Ramsey block ($\pi/2 \; – \; \pi \; – \; \pi/2$) inside the window, with a $\pi$ phase flip of the LO at the echo, cancels phase-blind terms and maps the probe quadrature $\hat X_\theta$ onto a small qubit $Z$-phase (red). 
}
  \label{fig:pulse_sequence}
\end{figure}

In the CM scheme, the coherent probe accumulates a phase shift
$\phi_b = \lambda \!\int_0^\tau n_a\,dt$ during its interaction with the thermal mode over the interaction time $\tau$. 
However, its photon number remains fixed, $n_b=|\alpha e^{i\phi_b}|^2=|\alpha|^2$, so a plain Ramsey evolution under the dispersive coupling $\chi_b \hat{n}_b \hat{\sigma}_z$ is phase blind.

To read out the probe’s phase fluctuations, we adopt a short, phase-sensitive mapping block~\cite{touzard2019gated,devoret2014quantum}. The thermal mode $a$ imprints a random phase $\phi_b$ on the probe $b$ during $\tau$; a following brief mapping then converts the probe's quadrature $X_\theta$ into a small qubit $Z$-phase. Shot-to-shot fluctuations of $\phi_b$ translate into fluctuations of the qubit phase and reduce the Ramsey visibility. From this visibility we infer $\mathrm{Var}(\phi_b)$, then $\bar n_a(T)$, and finally the temperature $T$.

During the short sensing window of duration $t_s$ between the two $\pi/2$ pulses, we apply a tiny local-oscillator (LO) displacement on the probe, $\beta=|\beta|e^{i\theta}$. The photon number seen by the qubit becomes
\[
n_b=\big|\alpha e^{i\phi_b}+\beta e^{i\theta}\big|^2
=|\alpha|^2+|\beta|^2+2|\alpha||\beta|\cos(\phi_b-\theta),
\]
which contains a phase-sensitive cross-term. With a Hahn echo inside the window and a $\pi$ flip of the LO phase ($\theta\!\to\!\theta+\pi$) at the echo, the phase-blind (number-like) contributions cancel while the quadrature term adds. The qubit therefore acquires a small, phase-sensitive angle
\[
\theta_q(\phi_b) = g_m\,2|\alpha|\cos(\phi_b-\theta),
\qquad
g_m = 2|\beta|\,\chi_b\,\tau_s.
\]
The net effect is an effective, brief longitudinal interaction that maps the probe quadrature onto a small qubit $Z$-phase. The mapping strength is set by a single, experimentally convenient gain knob $g_m$, and the analyzed quadrature is set by the LO phase~$\theta$ (typically biased near the maximum-slope point, $\phi_b-\theta\!\approx\!\pi/2$).

Operationally, the sequence is: (i) let the probe interact with the thermal mode for a time $\tau$ while the qubit is idle, so the probe acquires a random phase $\phi_b(T)$; (ii) turn on the mapping window $t_R$ bracketed by $\pi/2$ pulses, with the qubit–probe dispersive coupling active and a weak LO on the probe; (iii) insert a Hahn echo on the qubit and flip the LO phase by $\pi$ midway through the window. The echo removes phase-blind terms, while the LO phase flip preserves the desired quadrature term. The qubit then accumulates a tiny, phase-sensitive angle proportional to the probe quadrature at phase~$\theta$. Fig.~\ref{fig:pulse_sequence} illustrates a pulse sequence for this phase-to-quadrature mapping scheme.

Two practical remarks follow. First, the mapping is kept in the ``small-angle’’ regime: the LO injects $\ll 1$ photon on average and the window is much shorter than $T_2$, so backaction and higher-order corrections are negligible. Second, the same idea can be realized without an explicit LO by parametrically modulating the qubit (or a tunable coupler) at the probe frequency to generate a direct longitudinal coupling to $X_\theta$ during $t_s$~\cite{didier2015fast, blais2021circuit}.

\bibliography{CMQT}
\end{document}